\documentstyle[prd,aps,preprint]{revtex}
\begin{document}
%
\def\pr#1#2#3{ {\sl Phys. Rev.\/} {\bf#1}, #2 (#3)}
\def\prl#1#2#3{ {\sl Phys. Rev. Lett.\/} {\bf#1}, #2 (#3)}
\def\np#1#2#3{ {\sl Nucl. Phys.\/} {\bf B#1}, #2 (#3)}
\def\cmp#1#2#3{ {\sl Comm. Math. Phys.\/} {\bf#1}, #2 (#3)}
\def\pl#1#2#3{ {\sl Phys. Lett.\/} {\bf#1}, #2 (#3)}
\def\apj#1#2#3{ {\sl Ap. J.\/} {\bf#1}, #2 (#3)}
\def\ap#1#2#3{ {\sl Ann. Phy.\/} {\bf#1}, #2 (#3)}
\def\nc#1#2#3{ {\sl Nuo. Cim.\/} {\bf#1}, #2 (#3)}
\newcommand{\beq}{\begin{equation}}
\newcommand{\eeq}{\end{equation}}
\newcommand{\bea}{\begin{eqnarray}}
\newcommand{\eea}{\end{eqnarray}}
\newcommand{\aprime}[1]{#1^\prime}
\newcommand{\daprime}[1]{#1^{\prime\prime}}
\newcommand{\aop}{{A'_0}}
\newcommand{\nop}{{N'_0}}
\newcommand{\da}{\delta a}
\newcommand{\dn}{\delta n}
\newcommand{\dap}{\delta a^\prime}
\newcommand{\dapp}{\delta a^{\prime\prime}}
\newcommand{\dnp}{\delta n^\prime}
\newcommand{\dnpp}{\delta n^{\prime\prime}}
\newcommand{\db}{\delta b}
\newcommand{\dbp}{\delta b^\prime}
\newcommand{\Vdb}{{V(\delta b+\bar b\delta\phi)}}
\newcommand{\ie}{{\it i.e.}}
\newcommand{\eg}{{\it e.g.,}}
\draft
\preprint{\vbox{\hbox{\textsl{Dedicated to the memory of Detlef Nolte}}%
\hbox{~~}%
\hbox{\hskip1in UCSD/PTH 01--11}}}
\title{Bulk Observers in
Non-Factorizable Geometries}
\author{Benjam\'\i{}n Grinstein$^*$ and 
\underline{Detlef R. Nolte}$^\dagger$}
\address{$^*$Department of Physics,  University of California at San Diego,
         La Jolla, CA 92093 \\
         $^\dagger$Institute for Advanced Study, Princeton, NJ 08540}
\date{June 18, 2001}

\maketitle
\begin{abstract}
We consider five dimensional non-factorizable geometries where the
transverse dimension is bounded and the remaining (parallel)
dimensions are not. We study the construction of effective theories at
distances much longer than the transverse size. An observer unable to
resolve the transverse direction can only measure distances along the
parallel dimensions, but the non-factorizable geometry makes the length
of a curve along the parallel dimension sensitive to where on the
transverse direction the curve lies. We show that long geodesics that
differ in their endpoints only by shifts along the transverse
direction all have the same length to within the observer's
resolution. We argue that this is the correct notion of distance in
the effective theory for a bulk observer. This allows us to present 
a consistent interpretation of what is measured by observers that live 
either on a brane or in the bulk.
\end{abstract}

\narrowtext
\paragraph*{Introduction and Conclusions.}
\label{intro}
Randall and Sundrum have proposed a solution to the hierarchy problem 
based on a $Z_2$
five-dimensional orbifold with 3-branes at the fixed
points\cite{Randall:1999ee}. There is a negative cosmological constant
$\Lambda$ and tension on the branes, $V_{{\rm hid}}$ and $V_{{\rm
vis}}$. Then the metric
\beq
\label{eq:RSmetric}
ds^2=G_{MN}dx^Mdx^N=a^2(y)\eta_{\mu\nu}dx^\mu dx^\nu - dy^2 
\eeq 
solves Einstein's equations provided $a(y)= e^{-k|y|} $ and
\beq 
V_{hid}=-V_{vis}=24M^3 k
\quad \Lambda=-24M^3k^2
\eeq 
where $M$ is the fundamental 5-dimensional gravitational mass
scale. The brane with negative tension, $V_{vis}$, contains the
visible universe and is located at $y=y_c$ while the hidden brane,
with tension $V_{hid}$, is at $y=0$.  We will refer to this as the RS
model, to the $y$-coordinate as the transverse direction and to the
remaining ones as the parallel directions, since they are parallel to
the bounding branes.

In situations of physical interest the transverse direction is
microscopic. Macroscopic observers cannot resolve lengths as small as
the transverse dimension. Effectively they see a world of four
dimensions. Naively one would attempt to describe physics at long
distances by ignoring the transverse direction. This is, after all,
what is commonly done to infer the spectrum of Kaluza-Klein
excitations in factorizable geometries.
But for non-factorizable geometries the invariant interval of a line 
between the points
$(0,y_1)$ and $(\Delta x,y_2)$  depends on the transverse direction
$y$. This dependence can give results for the length that vary up to a
factor of $\exp(ky_c)$. The macroscopic observer is, however, unable
to discern the value of $y$. We are faced with the question: between
two points $0$ and $\Delta x$ what is the length measured by the
macroscopic observer? 
Many physical questions require an answer to this. For example, what does
the argument of a field, $\phi(x)$, in an effective theory represent?
Since mass is the inverse length for exponential fall-off of a static
field, what is the mass of a field? Although much of the
physics of these models has been understood\cite{Csaki:2000pp}, most of 
the research has been focused on the construction of effective field 
theories for brane observers and not bulk observers.
For an observer restricted to a brane the answer is well known. Distances 
measured by an apparatus made solely of components restricted
to a brane correspond to distances inferred by the induced metric on
the brane. The RS model has been analyzed in \cite{Grinstein:2001ny},
\cite{Ozeki} by using only the induced metric.
But an observer living in the bulk will be spread out over the microscopic 
transverse direction and it is not obvious which $4-dim$ metric will give 
the correct physical distances. 
Our main result is as follows:
distances measured by an apparatus made of unrestricted components (bulk
fields) correspond to distances measured on the visible brane,
provided these distances are much larger than the brane
separation. Specifically, the distance between points $(0,y)$ and
$(\Delta x,y')$ is, to high accuracy, $\exp(-ky_c)\Delta x$ for any
$y$ and $y'$, provided $\exp(-ky_c)\Delta x\gg y_c$.

 Effective theories for brane observers at long distances have 
been constructed by ``integrating-out'' the transverse direction. 
Let us review the modification by Ozeki and Shimoyama\cite{Ozeki} of the computation of Goldberger and Wise 
\cite{Goldberger:1999wh} for the case of a scalar field in
the background of the geometry in Eq.~(\ref{eq:RSmetric}).  The action
\begin{equation}
\label{eq:GWaction}
S={1\over 2}\int d^4 x dy \sqrt{G} \left(G^{AB}\partial_A \Phi
\partial_B \Phi - M^2 \Phi^2\right)
\end{equation}
admits a decomposition is in terms of the modes
\beq
\Phi(x,y)=\sum_n \psi_n(x) \xi_n(y)
\eeq
normalized so that the effective action is written in terms of the induced 
metric $g_{ind}$ according to \cite{Ozeki}
\begin{equation}
\int_{-y_c}^{y_c} dy e^{-2k|y|} \xi_n(y) \xi_m(y)= 
\left\{ 
\begin{array}{ll}
e^{-2k|y_c|}  \delta_{nm} & \mbox{observer at $y_c$}\\
\delta_{nm} & \mbox{observer at $y_0$,} \end{array} \right.
\end{equation}
and which satisfy
\beq
\label{eq:GWeigen}
-\frac{d}{dy}\left(e^{-4k|y|}\frac{d\xi_n}{dy}\right)
+M^2 e^{-4k|y|}\xi_n =m_n^2 e^{-2k|y|} \xi_n,
\eeq
giving
\beq
\label{eq:effaction}
S={1\over 2}\sum_n \int d^4 x 
\sqrt{g_{ind}}\left[g_{ind}^{\mu\nu}\partial_\mu \psi_n
\partial_\nu \psi_n - m_n^2 \psi_n^2\right]. 
\eeq
Analysis of the solutions of these equations give the correct mass
parameters for brane observers. However, it is not obvious how to apply this 
procedure to a bulk observer. 
Because of the finite transverse width of the bulk observer, the notion of 
an induced metric is not well defined. We argue that in the RS model 
distances measured by a bulk observer are equal to distances measured 
by an observer on the visible
brane up to errors of order of the length of the extra dimension. Therefore, 
we suggest that the KK mass spectrum as seen by a bulk observer is also 
equal to the spectrum seen by an observer on the visible brane.

\paragraph*{Brane Observers and Bulk Observers.}
\label{sec:observers}

Effective theories at long distances are meant to accurately describe
the world as seen by an observer who has limited resolution. In a
non-factorizable geometry we must make a distinction between observers
that live on one brane or the other and observers that live in the
bulk. We will refer to observers whose measuring devices are
constructed solely of fields and particles constrained to a brane as
``brane observers,'' and to those whose measuring devices are built of
unconstrained fields and particles  as ``bulk observers.'' As we will
see there are some qualitative and quantitative differences in the
observations they make, so the distinction is important.

We imagine that either observer can make a construction, following
Einstein, of locally inertial frames with identical meter sticks and
synchronized clocks. To better understand the construction it will be
necessary to know how to describe freely falling observers and how
observers measure distances and time. The former entails finding 
geodesics while the latter requires finding the minimum distance
between points on a fixed time hypersurface, that is, geodesics of
spatial sections.

We write the geodesic equation in the RS model for $0<y<y_c$, and the
case of $-y_c<y<0$ can be obtained formally by replacing $k\to-k$:
\begin{eqnarray}
\frac{d^2x^\mu}{d\tau^2}-2k\frac{dy}{d\tau}\frac{dx^\mu}{d\tau}&=&0\nonumber\\
\label{eq:ygeodesic}
\frac{d^2y}{d\tau^2}-
ke^{-2ky}\eta_{\mu\nu}\frac{dx^\mu}{d\tau}\frac{dx^\nu}{d\tau}&=&0.
\end{eqnarray}
For $\eta_{\mu\nu}\frac{dx^\mu}{d\tau}\frac{dx^\nu}{d\tau}>0$ time-like
geodesics are\cite{Muck:2000bb}
\begin{eqnarray}
x^\mu&=&x_0^\mu+\frac{v^\mu_0}{kv_0^2}\tan(k\tau)\\
y&=&-\frac1{2k}\ln(v_0^2\cos^2(k\tau)),
\end{eqnarray}
where $v_0^2\equiv\eta_{\mu\nu}v^\mu_0v^\nu_0$. The constants of
integration are $v_0^\mu$ and $x_0^\mu$, and they satisfy
$v_0^2>0$. One can also shift $\tau\to\tau-\tau_0$, but we have chosen
$\tau_0=0$ so that $dy/d\tau=0$ at $\tau=0$, for convenience. The
solution for $y=y(\tau)$ should be interpreted bearing in mind the
periodicity and identification conditions that define the
orbifold. Eq.~(\ref{eq:ygeodesic}) implies that at the brane $y=y_c$
the velocity $dy/d\tau$ is continuous. Using the reflection symmetry
and periodicity this implies that, if we insist in describing  the motion
on the interval $[0,y_c]$, the particle bounces off the brane at
$y=y_c$ reversing its velocity $dy/d\tau$. The motion $y(\tau)$ is
periodic, with period $\frac2k\cos^{-1}(e^{-ky_c}/\sqrt{v_0^2})$.

In terms of the coordinate time $t=x^0$, these solutions
give the expected straight line uniform motion in the large three
spatial dimensions:
\begin{eqnarray}
\label{eq:vecxvst}
x^i&=&x_0^i+\frac{v^i_0}{v_0^0}(t-t_0)\\
y&=&\frac1{2k}\ln\left(\frac1{v_0^2}
+\frac{v_0^2}{v_0^{02}}k^2(t-t_0)^2\right).
\end{eqnarray}
As expected for motion in a
gravitational field, the particle necessarily has non-vanishing velocity
(except at $\tau=0$) and acceleration along the transverse (fifth)
dimension. We see that a classical observer is either extended out
from $y=y_c$ into the bulk, or is, at best, in a limiting sense,
constrained to $y=y_c$. 

For completeness we give the case of massless particles too. If
$\eta_{\mu\nu}\frac{dx^\mu}{d\tau}\frac{dx^\nu}{d\tau}=0$ then
$dy/d\tau$ must be constant. But then $ds^2/d\tau^2\ge0$ only if
$dy/d\tau=0$. So this describes only massless particles (light-like
geodesics). This is consistent with the observation that the
Klein-Gordon equation in the warped background admits $y$-independent
solutions only for massless particles. Light-like geodesics are given,
for $\eta_{\mu\nu}\frac{dx^\mu}{d\tau}\frac{dx^\nu}{d\tau}>0$, by
\begin{eqnarray}
x^\mu&=&x_0^\mu+\frac{v^\mu_0}{k^2v_0^2\tau}\\
y&=&-\frac1{k}\ln(\pm k\sqrt{v_0^2}\tau).
\end{eqnarray}

What is the physical distance between points
on this space? Consider two points on a fixed time hypersurface, say
$t=0$, separated by some large coordinate distance $\Delta x$. It
would seem that the physical distance depends on the transverse
coordinate $y$ of these points. If so, how can one build an effective
four dimensional theory at physical distances much larger than $y_c$? 
The answer, as we now show,  is that at large $\Delta x$, the physical
distance these points is $\exp(-ky_c)\Delta x$, {\it independent} of
the $y$-coordinates of these points. The correction to this statement
is of order $y_c$, so an observer without the resolution to observe
the fifth dimension is also oblivious to this correction.

On a fixed $t$ hypersurface the geodesic equations are
\bea
\frac{d^2x^i}{d\tau^2}-2k\frac{dy}{d\tau}\frac{dx^i}{d\tau}&=&0\\
\frac{d^2y}{d\tau^2}+
ke^{-2ky}\delta_{ij}\frac{dx^i}{d\tau}\frac{dx^j}{d\tau}&=&0,
\eea
with solutions
\bea
x^i(\tau)&=&x^i_0 +\frac{\beta^i}{k^2\beta^2}\tanh[\omega(\tau-\tau_0)]\\
y(\tau)&=&-\frac1k\ln(k\beta\cosh[\omega(\tau-\tau_0)]),
\eea
where $\beta\equiv\sqrt{\delta_{ij}\beta^i\beta^j}$ and $\tau$ is an affine
parameter, $\tau\in[0,1]$. The parameter $\omega$ is simply the
physical length $\ell$ in units of $k$ along the geodesic,
$\omega=k\ell$. Given initial and final points, the parameters of the
geodesic are determined by
\bea
k\beta&=&
\frac{|\tanh[\omega(1-\tau_0)]+\tanh[\omega\tau_0]|}{k|\Delta x|},\\
\omega(1-\tau_0) &=&
\mbox{arcsinh}\left(\frac{d_1^2-1+d_1^2/d_0^2}{2d_1}\right),\\
\omega\tau_0 &=&
\mbox{arcsinh}\left(\frac{d_0^2-1+d_0^2/d_1^2}{2d_0}\right),
\eea
where $|\Delta x|^2\equiv \delta_{ij}(x^i(1)-x^i(0))(x^j(1)-x^j(0))$
and the quantities $d_{0,1}$ are simply the lengths in units of $k$
between $x^i(0)$ and $x^i(1)$ along curves of fixed $y$,
\bea
d_0 &=& k |\Delta x|e^{-ky(0)},\\
d_1 &=& k |\Delta x|e^{-ky(1)}.
\eea
To understand this solution consider the case $y(0)\le y(1)$ which is
described by $\tau_0\ge1/2$. If $\tau_0\ge1$ the coordinate $y(\tau)$
increases monotonically from $y(0)$ to $y(1)$, while for
$1/2\le\tau_0<1$ the solution $y(\tau)$ increases monotonically from
$y(0)$ beyond $y(1)$ (at $\tau=2\tau_0-1$) to a maximum at $y(\tau_0)$
and decreases monotonically back to $y(1)$. The transition between
these two distinct behaviors occurs at $\tau_0=1$, that is, at a
critical $x$ separation given by
\beq
\label{eq:critlength}
k |\Delta x|_{\rm crit}e^{-ky(1)} = \sqrt{1-e^{-2k(y(1)-y(0))}}.
\eeq
For  $|\Delta x|<|\Delta x|_{\rm crit}$ the solution has $\tau_0>1$.

The solutions to the geodesic equation above do not describe the
minimum distance path between points in the orbifold of the RS model
because it neglects the presence of fixed points. At large separation
$|\Delta x|$ the solution above extends to the region $y>y_c$. The
shortest path is actually along a $\tau_0=1$ geodesic from $y(0)$ to
$y_c$, then along the brane $y=y_c$ and finally back from $y_c$ to
$y(1)$ along a second ($\tau_0=0$) geodesic. The shortest path between two
points has length
\beq
\label{eq:physlength}
\ell_{\rm phys}=|\Delta x|e^{-ky_c}+(2y_c-y(0)-y(1)+\ln4-2)+
{\cal O}(e^{-k(y_c-\min(y(0),y(1))}).
\eeq
This applies only provided the length along $x$ is large enough that
the above geodesic would hit the $y=y_c$ brane. The condition for this
is, from Eq.~(\ref{eq:critlength}), that the length measured along the
$y=y_c$ brane be larger than $1/k$. 

We can now understand what a brane and a bulk observer are, their
similarities and differences. First, Eq.~(\ref{eq:vecxvst}) tells us
that if we ignore the $y$-motion of a particle in the bulk the motion
in the transverse space is like that of a free particle in flat
space. Bulk observers, or rather their meter sticks, who have
resolution smaller than the brane separation $y_c$ are presumably much
bigger than $y_c$ themselves. They are made of many particles
distributed over the $y$-direction. A priori one could expect these
particles would spread in $x^i$ even if they have the same initial
velocity, but we see this is not the case (a fortunate state of
affairs for the observer).

Second, Eq.~(\ref{eq:physlength}) tells us that the meter stick this
bulk observer will use to measure distances between two points does not
depend on the $y$-coordinates of these points. The error introduced by
ignoring the endpoint of the meter stick is smaller than the distance
between the branes, but we already assumed that the resolution of this
observer was worse. 

Third, Eq.~(\ref{eq:physlength}) shows that the distance measured by
the bulk observer between two points is what a negative tension brane
observer (that is, an observer on the $y=y_c$ brane) would measure
(between the points projected onto the brane). The distance between
these points measured by the positive tension brane observer is,
however, exponentially larger.

Last, similar conclusions hold for measurement of time, since motion
is uniform in $x^i(t)$. Observers can measure two units of
time by reflecting a pulse of light off a mirror set at the end of a
meter stick.

\paragraph*{Interpretation Of The Effective Action.}
\label{sec:KK}
 
The meaning of effective theories with the transverse direction
integrated out is now clear. For a definite example consider the bulk scalar
action of~(\ref{eq:GWaction}) and the corresponding effective
theory~(\ref{eq:effaction}).  Instead of formally integrating out the
$y$-direction, we study the response of the bulk scalar to arbitrary
sources smeared over the resolution of the detecting apparatus and
separated by distances larger than the resolution. The Green function
of the Klein-Gordon equation, $\Delta(x,y;x',y')$ can be written in
terms of the four dimensional Green function for a particle of mass
$m_n$, $\Delta^{(4)}(x-x';m_n)$, as\cite{Grinstein:2001ny}
\bea
\label{eq:Deltaexpansion}
\Delta(x,y;x',y')&=&-\sum_n\int\frac{d^4q}{(2\pi)^4}
\frac{e^{iq\cdot(x-x')}}{q^2-m_n^2}R_n(y,y')\nonumber\\
&=&\sum_n\Delta^{(4)}(x-x';m_n)R_n(y,y').
\eea
Since the residue factorizes,
$R_n(y,y')=r_n(y)r_n(y')$, the full Green function $
\Delta(x,y;x',y')$ can be obtained in a four dimensional description
by coupling sources $j(x)$ and $j'(x)$ to the linear combinations
$\sum_n r_n(y)\psi_n(x)$ and $\sum_n r_n(y')\psi_n(x)$, respectively.
The spectrum, $m_n$, is determined as solutions to  
\beq
{\tilde N_\nu(m_nz_1)\tilde J_\nu(m_nz_2)
-\tilde J_\nu (m_nz_1)\tilde N_\nu (m_nz_2)}=0,  
\eeq
where $z_{1,2}=e^{ky}/k|_{y=0,y_c}$, $\nu=\sqrt{4+m^2/k^2}$,
and we have introduced the shorthand
$ \tilde Z_\nu(z) = (1-\frac\nu2)Z_\nu(z)+\frac{z}2Z_{\nu-1}(z)$ for
the combinations of Bessel functions of first and second kind, $J_\nu$
and $N_\nu$. This is precisely the same equation as found by Goldberger and Wise by
means of a different method\cite{Goldberger:1999wh}, namely, direct
diagonalization of the action integral. For small $m$, the low
excitation number spectrum has $m_n\sim a(y_c)k$. As above, the full
Green function can be written as a sum over poles,
Eq.~(\ref{eq:Deltaexpansion}), where the residues factorize.

By the arguments of the previous section, an observer on the visible
(negative tension) brane sees particles of physical mass
$m_n/a(y_c)\sim k$. Their `overlap', or wave-function on the visible
brane, is given by $r_n(y_c)$. Similarly, since long distances in the
bulk correspond to long distances on the visible brane, bulk observers
also see particles of physical masses and splittings of order $k$. It
is only observers on the hidden (positive tension) brane who see
exponentially suppressed masses. There is a simple physical
interpretation. Hidden brane observers see masses that have climbed up
a potential well and are therefore red-shifted precisely by the warp
factor.

The effective action for this theory for a bulk observer or for an
observable brane observer is just as in Eq.~(\ref{eq:effaction}), with
the subsidiary information that the coordinates $x^\mu$ do not measure
physical distance, but exponentially large distance. When the
coordinates are rescaled so that they actually correspond to physical
distance the effective action is of the same form, but the masses are
replaced $m_n\to\exp(ky_c)m_n$. Interactions can be included. They are
computed by considering $n$-point functions. Similarly, the effective
action in~(\ref{eq:effaction}) correctly accounts for the physics seen
by the hidden brane observer, since $x^\mu$ does correspond to
physical distance there.

\paragraph*{Generalizations} Similar results are obtained for a wide
class of non-factorizable geometries. A simple case is that in which
a space built on an $S_2/Z_2$ orbifold,
\beq
\label{eq:genmetric}
ds^2=e^{2A(y)}\eta_{\mu\nu}dx^\mu dx^\nu-dy^2,
\eeq
has a warp factor $A$ that is locally a minimum at both
fixed points. At large parallel separations the bulk observers see
distances as measured along the brane of smaller warp factor.

An interesting case is the metric for a space with a single brane at a
fixed point in Anti-de Sitter 5-space ($AdS_5$) with Anti-de Sitter
($AdS_4$) sections\cite{DeWolfe:2000cp}. The metric is as in
Eq.~(\ref{eq:genmetric}), but with $\eta$ replaced by the metric of
$AdS_4$. The warp factor is
\beq
A=\log\left(\cosh{c-|y|\over L}{\Big /}\cosh{c\over L}\right),
\eeq
decreasing from the brane to $y=c$ where it has a minimum and then
growing again. Fixed-$t$ geodesics at large parallel separation
$\Delta x$ go mostly very close to the hypersurface $y=c$. There is no
brane located there, still distance scales for a bulk observer are as
if measured along a $y=c$ brane. The graviton is localized on the
brane at $y=0$\cite{Karch:2001ct}. The spectrum of massive excitations
of the graviton is observed to be exponentially lighter by a brane
observer than by a bulk observer, giving an amusing inverted
hierarchy.

\bigskip

{\it Acknowledgments} We would like to thank Ken Intriligator, Ira
Rothstein and Witold Skiba for discussions.  This work is supported by
the Department of Energy under contract No.\ DOE-FG03-97ER40546.



\end{document}